\documentclass[aps,prd,showpacs,twocolumn]{revtex4}

\usepackage{epsfig}

\newcommand{\geqnew}{\stackrel{>}{\!\ _{\sim}}}
\newcommand{\leqnew}{\stackrel{<}{\!\ _{\sim}}}

\begin{document}
\begin{flushleft}
TRI-PP-03-05 \\
ADP-03-121/T559 \\
\end{flushleft}

\title{Measurement of $\tan\beta$ in associated $t H^\pm$ Production 
in $\gamma\gamma$ Collisions}
 \author{M. A. Doncheski$^a$, Stephen Godfrey$^{b,c,d}$ and Shouhua 
Zhu$^b$}
\affiliation{$^a$Department of Physics, Pennsylvania State University, \\
Mont Alto, PA 17237 USA \\ 
$^b$Ottawa-Carleton Institute for Physics \\
Department of Physics, Carleton University, Ottawa, Canada K1S 5B6 \\
$^c$Special Research Centre for the Subtatomic Structure of Matter\\
University of Adelaide, Adelaide South Australia 5000, Australia \\
$^d$ TRIUMF, 4004 Wesbrook Mall, Vancouver B.C. Canada V6T 2A3}

\date{\today}

\begin{abstract}
The ratio of neutral Higgs field vacuum expectation values, 
$\tan\beta$, is one of the most important parameters to determine in 
type-II Two-Higgs Doublet Models (2HDM), specifically the Minimal 
Supersymmetric Standard Model (MSSM). Assuming the energies and 
integrated luminosity of a future high energy $e^+e^-$ linear collider
of $\sqrt{s}=500$, 800, 1000, and 1500~GeV and ${\cal L}=1$~ab$^{-1}$ 
we show that associated $t H^\pm$ 
production in $\gamma\gamma$ collisions
can be used to make an accurate determination of 
$\tan\beta$ for low and high $\tan\beta$ by precision measurements of 
the $\gamma\gamma \to H^\pm t +X $ cross section.  
\end{abstract}

\maketitle

\section{Introduction}

A fundamental open question of the standard model (SM) 
is the origin of electroweak symmetry breaking 
(EWSB)\cite{dawson99,carena02}.  
The simplest description of EWSB results in one neutral scalar 
particle which, however, has well known problems associated with it.  
A priori, a more 
complicated Higgs sector is phenomenologically just as viable 
\cite{gunion02a}.  The next 
simplest case is the general two Higgs doublet model (2HDM).  
A constrained 
version of the 2HDM is a part of the minimal supersymmetric extension of the 
SM (MSSM) \cite{susy,mssm} 
where spontaneous symmetry breaking is induced by two 
complex Higgs doublets and leads to five physical scalars; the neutral CP-even 
$h^0$ and $H^0$ bosons, the neutral CP-odd $A^0$ boson, and the charged 
$H^\pm$ bosons.  At tree level the MSSM Higgs sector has two free parameters 
which are usually taken to be the ratio  of the vacuum expectation values of 
the two Higgs doublets, $\tan\beta=v_2/v_1$, 
where $v_2$ couples to the up-type quarks and $v_1$ to the down-type 
quarks, and the mass of the $A^0$ boson, $m_A$.  
The elucidation of  EWSB is the primary goal of the Large Hadron 
Collider at CERN (LHC) and the proposed high energy $e^+e^-$ Linear 
Collider (LC) 
\cite{teslatdr,higgs,chargedhiggs,komamiya88,chao93,kanemura,moretti02,gunion97}. 

The ratio of neutral Higgs field vacuum expectation values, 
$\tan\beta$, is a key parameter needed to be determined in 
type-II Two-Higgs Doublet Models and the MSSM. 
In addition to providing information 
about the structure of the non-minimal Higgs sector, the measurement 
of this parameter also provides an important check of SUSY structure 
as this parameter also enters the chargino, neutralino, and third 
generation squark matrices and couplings \cite{sparticles}.  
Nevertheless, the Yukawa 
couplings of Higgs bosons are the most direct way to probe the 
structure of the vacuum state of these theories. 
To address this issue 
a number of recent studies have examined how one could measure 
$\tan\beta$ at the LHC \cite{lhc} and 
future high energy $e^+e^-$ colliders
\cite{he02,djouadi92,gunion02,Boos:2003vf,barger01,feng97}.
A $\gamma\gamma$ 
``Compton Collider'' option, from backscattered laser light off of 
highly energetic electron beams, has been advocated as a valuable part 
of the linear collider program  which could make 
crucial measurements of the Higgs sector \cite{telnov}.
If charged Higgs 
bosons were observed, cross section measurements could potentially give 
information about the underlying theory.
Analysis of the process $e^+e^-\to H^+H^-$ indicates that the absolute 
event rate and ratios of branching ratios in various $H^+H^-$ final 
state channels will allow a relatively accurate determination of 
$\tan\beta$ at low $\tan\beta$ \cite{gunion97}.
The branching ratios for $H$, $A$, and $H^\pm$ are sensitive to 
$\tan\beta$ when $\tan\beta$ is less than roughly 20 so that a precise 
measurements of branching ratios can give a good determination of 
$\tan\beta$.  

Another possibility for measuring $\tan\beta$ is associated $tH^\pm$ 
production in $\gamma\gamma$ collisions \cite{doncheski03}.  The subprocess 
$b\gamma \to H^- t$ proceeds via  $b\gamma$ fusion and utilizes the $b$-quark 
content of the photon.  Despite the fact that there is good agreement 
on the $b$-quark content of the photon 
between the existing sets of photon 
parton distribution functions, at this time 
there are no experimental data to back up the theoretical calculations.
This is not an unsurmountable problem.  A direct measurement of the 
$b$-quark content of the photon is possible in a linear $e^+e^-$ or $e\gamma$ 
collider  (see for example Ref. \cite{doncheski97}) 
and such a measurement will need 
to be made before, and independently of, the process discussed in this 
paper.  In this report we study how well the subprocess 
$b\gamma \to H^- t$ can be used to measure $\tan\beta$.  We find that it can 
be used to make a good determination of $\tan\beta$ for most of the parameter 
space with the exception of the region around $\tan\beta\simeq 7$.  As such, 
it is a useful complement to other measurements for studying $\tan\beta$ 
\cite{gunion02}.

\section{Calculations and Results}

The process $\gamma\gamma\to tH^\pm +X$ 
makes use of the $tbH^\pm$ interaction to measure 
$\tan\beta$.  The interaction is given by \cite{HHguide,barger}:
\begin{equation}
i {{V_{tb}}\over {\sqrt{2}v}} [m_b \tan\beta (1+\gamma_5) + m_t \cot 
\beta (1-\gamma_5)]
\end{equation}

\begin{figure}[t]
\begin{center}
\centerline{\epsfig{file=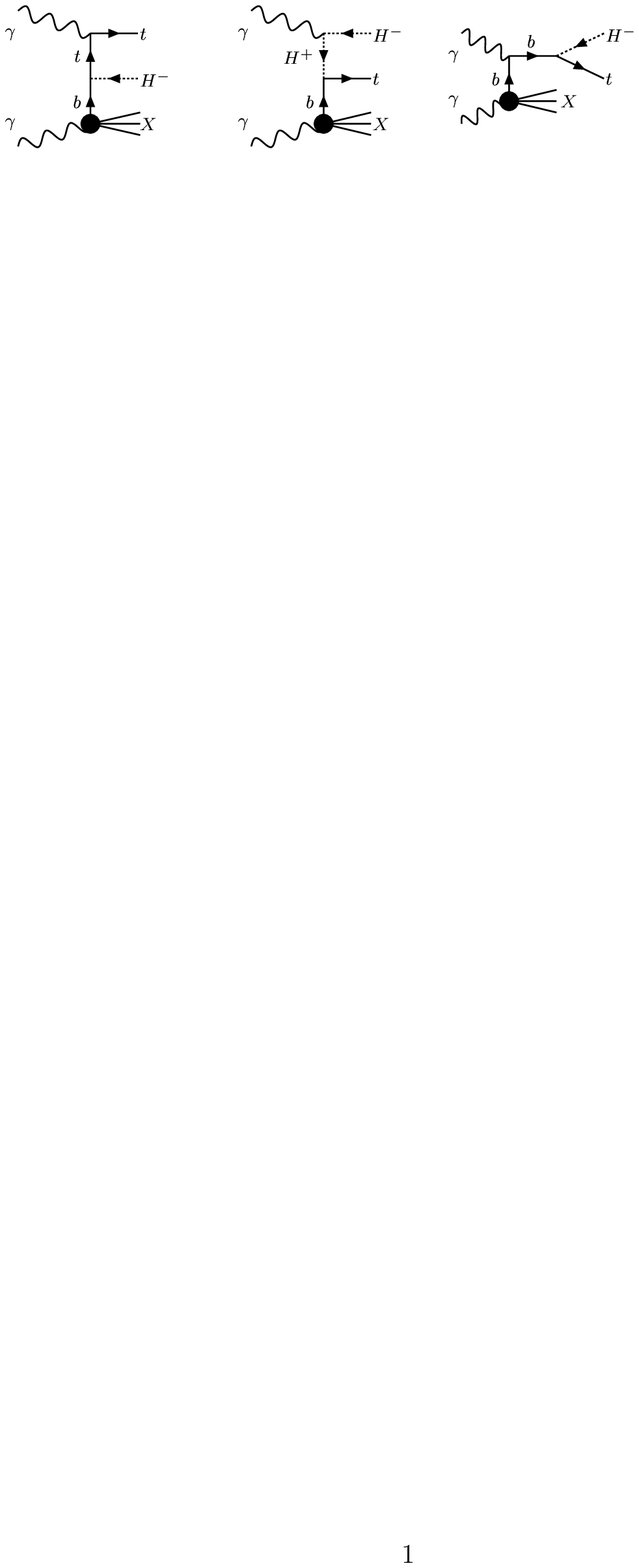,width=3.5in,clip=}}
\end{center}
\caption{Feynman diagrams for the process 
$\gamma\gamma\to t H^- + X$ proceeding via the 
$b$-quark content of the photon.}
\label{Fig1}
\end{figure}

To calculate the process $\gamma\gamma\to H^- t +X $ we calculate the 
subprocess $\gamma b \to H^- t$ and convolute it with the $b$ 
distribution in the photon.  This is illustrated in Fig. 1.
It has been shown that this is a 
reasonable approximation for the energies and kinematic regions 
we are considering in this paper \cite{doncheski97}.  The amplitude 
squared for the subprocess $b\gamma\to t H^-$ is given by: 
\begin{widetext}
\begin{eqnarray}
\sum |M(b\gamma &\to& t H^\pm) |^2  =   
8\sqrt{2}G_F \pi\alpha (m_b^2 \tan^2\beta + m_t^2/\tan^2\beta) \nonumber\\
& & \left\{ {
{{Q_b^2}\over{s^2}} 2 p_t \cdot p_\gamma p_b\cdot p_\gamma 
-{{Q_t^2}\over{(t-m_t^2)^2}} 2[m_t^2 ( p_b \cdot p_t -p_b \cdot 
p_\gamma ) - p_t \cdot p_\gamma p_b \cdot p_\gamma ] } \right. \nonumber\\
& & 
- {{2}\over{(u-m_t^2)^2}}  p_t\cdot p_b (m_t^2 -2 p_b \cdot p_t ) 
-2 {{Q_b Q_t}\over {s(u-M_h^2)}} [2(p_t\cdot p_b + p_t \cdot p_\gamma)
(p_t\cdot p_b - p_b \cdot p_\gamma) - m_t^2 p_b \cdot p_\gamma ] \nonumber\\
& & -2 {{Q_b}\over{s(u-M_h^2)}} 
[2 p_t \cdot p_b (p_b\cdot p_t + p_t \cdot p_\gamma ) 
- m_t^2 p_b \cdot p_\gamma ] \nonumber\\
& & + \left. {  {{Q_t}\over{(t-m_t^2)(u-M_h^2)}} [2 p_t \cdot p_b (p_t \cdot p_b - 
p_b \cdot p_\gamma) + m_t^2 (p_b \cdot p_\gamma -2 p_t \cdot p_b ) ] }\right\}
\end{eqnarray}
\end{widetext}
In our numerical results we include an additional factor of 2 
from producing either an $H^+$ or $H^-$ and a factor of 2 
because the $b$-quark can come from either initial photon. 
We note that we are ignoring $m_b$ compared to $m_t$ and $M_H$ 
everywhere except in the coupling factor where $m_b \tan\beta$ can be 
comparable to $m_t/\tan\beta$.  
To obtain the subprocess cross section 
we integrate the matrix element using Monte Carlo integration 
\cite{barger} and 
convolute the subprocess with the $b-$quark distribution in the photon 
and the photon spectrum,  either the energy distribution obtained from 
backscattering a laser from an electron beam \cite{backlaser} or the 
Weizs\"{a}cker Williams distribution \cite{WW}.
The cross sections are found by evaluating the following expression:
\begin{equation}
\label{cross}
\sigma = \int dx_1 dx_2 dx_3 \;
f_{\gamma/e}(x_1) f_{\gamma/e}(x_2) f_{b/\gamma}(x_3)
\; \hat{\sigma}(\hat{s})
\end{equation}
where $\hat{\sigma}(\hat{s})$ is the subprocess 
cross section for C of M energy $\sqrt{\hat{s}}$.
We have only
included tree-level contributions and are aware that higher order
corrections are 
likely to be non-negligible.  Nevertheless, we feel that our approach is 
satisfactory for a preliminary study to gauge the potential of this 
process for measuring $\tan\beta$.  
Our calculations have explicit dependence on the $b$ and $t$-quark masses.  
We take $m_b=4.4$~GeV, $m_t=175$~GeV, and $V_{tb} \sim 1$.  
In addition, we used 
$M_W=80.41$~GeV,  $G_F=1.166\times 10^{-5}$~GeV$^{-2}$, 
and $\alpha=1.0/128.0$ \cite{pdg}.  
To obtain numerical results we used the Gl\"uck Reya Vogt (GRV) 
distributions \cite{grv}.  Other distributions are available \cite{pds1,pds2} 
and with recent LEP and HERA data, updates are forthcoming 
\cite{cornet02}.
The cross section is shown in Fig 2
as a function of $M_H$ for $\sqrt{s}=500$~GeV for $\tan\beta=1.5$, 3, 
7, 30, and 40.

\begin{figure}[t]
\begin{center}
\centerline{\epsfig{file=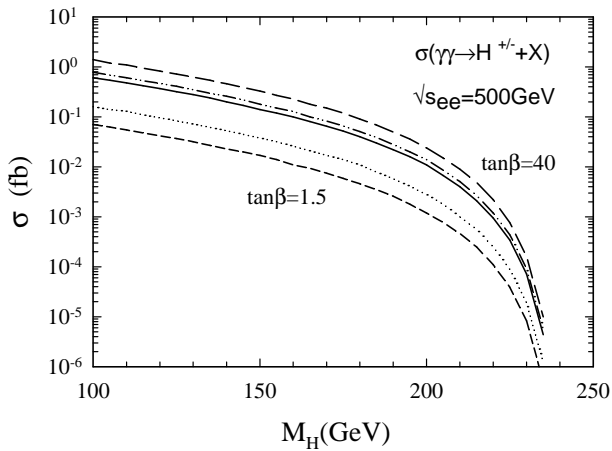,width=3.2in,clip=}}
\end{center}
\caption{$\sigma(\gamma\gamma\to t H^- +X)$ for the $\gamma\gamma$
backscattered laser case with $\sqrt{s}_{ee}=500$~GeV and 
$\tan\beta=1.5$ (short-dashed line), 3 (dotted line), 7 (solid line),
30 (dot-dot-dashed line), and 40 (long-dashed line).}
\label{Fig2}
\end{figure}

The process we are studying has two initial state photons.  In 
addition to the backscattered laser photons the initial state photons
can also be 
Weiszacker-Williams photons bremstrahlung from the initial electron 
beams so that in addition to $\gamma\gamma$ collisions we also 
consider $e\gamma$ and $e^+e^-$ collisions.  We show the cross 
sections for the $\gamma\gamma$, $e\gamma$ and $e^+e^-$ modes in Fig 3 
for $\tan\beta=3$ and 40.  One sees that the cross sections decrease 
by about an order of magnitude for each replacement of backscattered 
laser photons with Weizacker-Williams photons although the cross 
sections for backscattered laser photons give lower kinematic limits 
due to the cutoff in their energy spectrum.  Because good 
statistics are central to extracting precision measurements of 
$\tan\beta$ we find that the error increases substantially for the 
$e\gamma$ and $e^+e^-$ cases compared to the $\gamma\gamma$ case.

\begin{figure}[t]
\begin{center}
\centerline{\epsfig{file=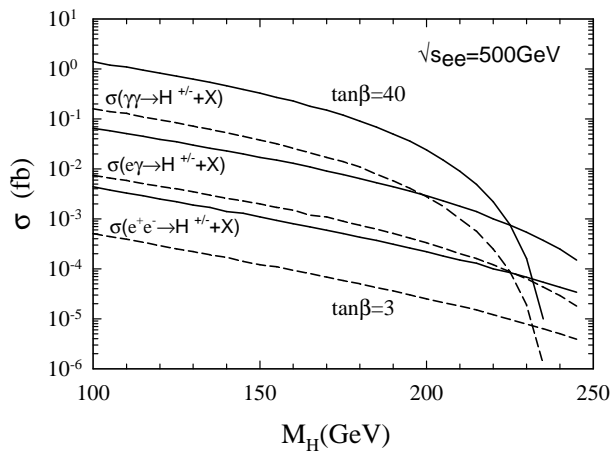,width=3.2in,clip=}}
\end{center}
\caption{$\sigma(\gamma\gamma\to t H^- +X)$ for the $\gamma\gamma$, 
$e\gamma$ and $e^+e^-$ cases  with $\sqrt{s}_{ee}=500$~GeV and for
$\tan\beta=3$ (dashed lines) and 40 (solid lines).  For each set the 
largest cross section is for the $\gamma\gamma$ case, the middle cross 
section for the $e\gamma$ case and the smallest cross section for the 
$e^+e^-$ case.}
\label{Fig3}
\end{figure}

This is a preliminary analysis; as such we do not explicitely
consider detector efficiencies nor include backgrounds.  
They are taken into account by assuming different values for the 
overall detection efficiency which we discuss below.
Additionally, we don't include decay branching ratios.  The 
branching ratios are well known \cite{higgsbr,mssmdecays}, and dominated by 
$tb$ or $\tau \nu_\tau$ depending on the $H^\pm$ mass and $\tan \beta$.  

A precise measurement of $\tan\beta$ depends on a precise measurement of the 
$t H^\pm$ production cross section.  Thus, the issue of extracting the event 
signal from backgrounds and the efficiency with which this can be accomplished 
becomes the determining factor in the $\tan\beta$ measurement.  The most 
significant backgrounds which need to be separated out are 
$\gamma\gamma\to t\bar{t}$ and $\gamma\gamma\to t\bar{b}+\bar{t}b$ (where one 
of the $b$'s is lost down the beam) and $\gamma\gamma \to t\bar{b} + W^-$ 
(where the $b$ is lost down the beam and the $W$ decays to 
$\tau+\bar{\nu}_\tau$).  The final state, and thus the most significant 
background, depends on the mass, $M_{H^\pm}$.  If $M_{H^\pm}>m_t$, 
$H^\pm \to tb$ will be the dominant decay mode and if $M_{H^\pm}<m_t$, 
$H^\pm \to \tau\nu$ will be the dominant decay mode.  In either case, these 
backgrounds can be distinguished from the signal by reconstructing the final 
states.

For the case that $M_{H^\pm} > m_t$ there will be two $t$-quarks in 
the final state.  
Each of the $t$-quarks will decay 
to $bW$ with the $W$'s subsequently decaying 
to either $q\bar{q}$ ($jj$) or $\ell \nu$ final states.
The hadronic 
$W$ decays lead to three jets in the final state while the leptonic 
$W$ decay leads to one jet, an isolated lepton and missing energy. 
To reconstruct the event one needs to associate the final state jets, 
leptons, and missing energy with the originating parton. 
$b$-tagging is crucial to reconstructing the event and there is a 
tradeoff of efficiency versus mistagging a $c$ or $uds$ jet.  
The $W$ is reconstructed from the non-$b$-tagged 
jets and the $t$ is reconstructed through a constrained fit to the $W$ 
and the $b$-jet that gives the best fit to the $t$-quark mass.  For 
leptonic $W$ decay the longitudinal component of the neutrino momentun 
can be fixed using the $W$ mass constraint with the missing transverse 
momentum.  The remaining $b$-jet is paired with the reconstructed top 
to reconstuct the $H^\pm$.  As it is not possible to know which of the 
two top quarks originates from the Higgs decay there is a large 
combinatorial background from signal events and the associated
difficulty in assigning 
non-$b$ jets to the correct top cluster.  However, the constrained 
kinematic fits can be used to distinguish between the signal and 
background.
After imposing selection criteria signal events can be 
selected with some efficiency but with some fraction of the events 
misidentified background.  

For the case $M_{H^\pm} < m_t$ the final state will consist of a charged 
$\tau$-lepton, a $\tau$-neutrino and a $t$-quark, which will lead to 0, 1 or 2 
charged leptons, missing $E_{_T}$ and at least one jet.  As above, the 
$t$-quark can be reconstructed from the $b$-tagged jet and hadronic or 
leptonic $W$ decay products, with some efficiency and some probability of 
mis-identification.  Given the success of the LEP collaborations in measuring 
$\tau$ properties, for example the $\tau$ lifetime \cite{lifetime_tau} and 
polarization \cite{pol_tau_ALEPH,pol_tau_DELPHI,pol_tau_L3,pol_tau_OPAL}, 
$\tau$-leptons can be reconstructed successfully in a collider environment.  
The $\tau$-lepton can be reconstructed from low multiplicity, collimated 
(one-prong) hadronic decays, with the $\tau$ mass constraint determining the 
transverse momentum of the $\nu_\tau$ produced in the $\tau$ decay.  Such 
single prong decays account for approximately 46\% of all $\tau$ decays 
($Br(\tau \rightarrow \pi \nu) \sim 12\%$, 
$Br(\tau \rightarrow \rho \nu) \sim 26\%$ 
and $Br(\tau \rightarrow a_1 \nu) \sim 8\%$).  As noted above, LEP 
collaborations have been successful in reconstrucing $\tau$ events; for 
example, ALEPH \cite{pol_tau_ALEPH} report $60\% - 80\%$ efficiency in 
reconstrucing single prong $\tau$ decays, with mis-identification generally 
between $4\% - 20\%$.  ALEPH report on two $\tau$ identification techniques, 
one with higher efficiency but higher mis-identification probability than the 
other.  This is a generic property of $\tau$ reconstruction.  Other LEP 
collaborations report similar efficiencies and sample purities 
\cite{pol_tau_DELPHI,pol_tau_L3,pol_tau_OPAL}.  Leptonic $\tau$ decays, on 
the other hand, have two neutrinos, each carrying transverse momentum, making 
the full reconstruction of the $\tau$ problematic.  Furthermore, the 
measurement considered here is going to be made after the discovery and 
measurement of the charged Higgs boson, so the Higgs boson mass can be used to 
constrain the transverse momentum of the $\nu_\tau$ produced in the charged 
Higgs boson decay.  We expect, then, that using only single prong, hadronic 
$\tau$ decay modes that $\tau$-leptons will be reconstructed with approximatly 
30\% efficiency.  However, to be conservative, we assume similar 
reconstruction efficiencies for the $t$ and $\tau$, and give sensitivity 
curves for a common set of assumed reconstruction efficiencies.  A more 
complete analysis must include detector efficiency, backgrounds and branching 
ratios.  

How well the signal can be separated from the background
depends critically on the $b$ 
tagging efficiency which in turn depends details of the detector.  
Studies using detector simulations have been performed on 
related processes.  They have found that the efficiency is inversely 
related to the purity of the $t$-quark sample.  {\it i.e.}, the higher the 
purity the smaller the efficiency.  In practice one wants to find the 
point on this efficiency-purity curve that maximizes the signal to 
background ratio.  Detailed studies on processes similar to the one we 
are studying use $b$-tagging efficiency $\epsilon \simeq 10\%$ 
for a high purity ($\geqnew 90\%$ purity), yet statistically significant 
sample.  Given that it is 
impossible to know the exact value to use for this important parameter 
without a real detector with measured properties we give results 
for different values of the event reconstruction 
efficiency.  With this approach, experimentalists can decide which 
value most closely corresponds to their detector.

Another consideration is the variation of $BR (H^\pm \to tb)$ with 
$\tan\beta$ and $M_{H^\pm}$, especially for small valued $\tan\beta$ where 
the $tb$ mode competes with the $\tau\nu$ mode.  As noted above, we assume 
similar reconstruction efficiencies for these two modes and assume that 
experiments can measure this process through both charged Higgs boson decay 
modes.  

To obtain our results we assume 1~ab$^{-1}$ for the integrated 
luminosities,  the standard used in many  LC studies, to 
estimate event rates and hence statistical errors.  
We consider 4 centre of 
mass energies appropriate to the various linear collider energies 
proposed: $\sqrt{s}=500$, 800, 1000, and 1500~GeV.

The upper and lower limits on $\tan\beta$ are obtained as the values of
$\tan\beta \pm \Delta\tan\beta$ for which the cross sections are 
statistically consistent 
with the cross sections obtained using the 
central value of $\tan\beta$ given on the x-axis.  In Fig 4 we show 
the cross section {\it vs.} $\tan\beta$ along with the 1-$\sigma$ errors on 
$\tan\beta$ presented as the ratio
$\Delta\tan\beta/\tan\beta$.  Over most of 
the $\tan\beta$ range there is a strong dependence of the cross 
section on $\tan\beta$ so that a precise cross section measurement 
will result in a good determination of $\tan\beta$.  However, 
at $\tan\beta \simeq 6-7$ the cross section is at a minimum so that 
$\Delta\sigma/\Delta\tan\beta \sim 0$ resulting in very little sensitivity 
to $\tan\beta$ in this region.  The minimum of cross section can be seen 
both numerically and analytically.  The matrix element squared, given in 
Eqn.~2, is proportional to $m_b^2 \tan^2 \beta + m_t^2/\tan^2 \beta$; 
minimizing this with respect to $\tan \beta$, we find 
$\tan \beta_{min} = \sqrt{m_t/m_b} \sim 6.3$ for the values of the top and 
bottom masses used.  The actual minimum of sensitivity to $\tan \beta$ will 
depend on the detailed behavour of the cross section, but it is no surprise 
that this process is insensitive to the value of $\tan \beta \sim 6-7$.
The assymetry in the sensitivity curves, which is more apparent in 
subsequent figures, is the simple consequence that when 
$\tan\beta \leqnew 6$, as $\tan\beta$ is increased the cross section 
enters the region least sensitive to $\tan\beta$ while for 
$\tan\beta \geqnew 6$, when $\tan\beta$ is decreased the cross section 
enters this region.

\begin{figure}
\begin{center}
\centerline{\epsfig{file=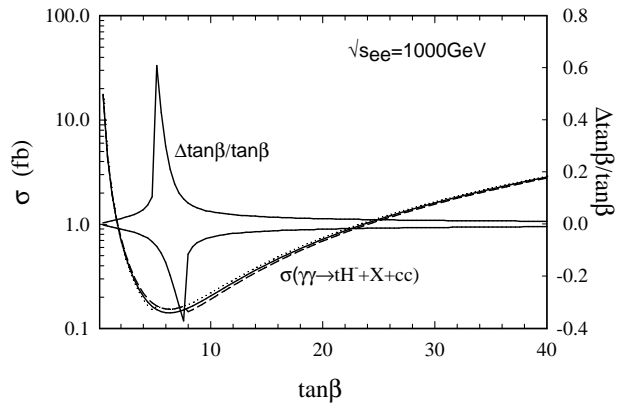,width=3.2in,clip=}}
\end{center}
\caption{$\sigma(\gamma\gamma\to t H^- +X)$ {\it vs.} $\tan\beta$ for the 
backscattered laser case and the sensitivities to $\tan\beta$ based 
only on statistical errors (solid lines) for $\sqrt{s}_{ee}=1$~TeV and 
$M_H = 200$~GeV.  For the cross sections, the solid 
line represents the expected cross section at the nominal value of 
$\tan\beta$, while the dashed (dotted) line represents the expected cross 
section at $\tan\beta - \Delta\tan\beta$ ($\tan\beta + \Delta\tan\beta$).
}
\label{Fig4}
\end{figure}

We have previously pointed out that the measurement of $\tan\beta$ is 
much weaker in the $e\gamma$ and $e^+e^-$ modes of a future linear 
collider.  In Fig 5 we show $\Delta\tan\beta/\tan\beta$ for the 
$\gamma\gamma$, $e\gamma$ and $e^+e^-$ modes for 
$\sqrt{s}_{ee}=1$~TeV and $M_H=250$~GeV.  We will henceforth only show 
results for the $\gamma\gamma$ case.

\begin{figure}
\begin{center}
\centerline{\epsfig{file=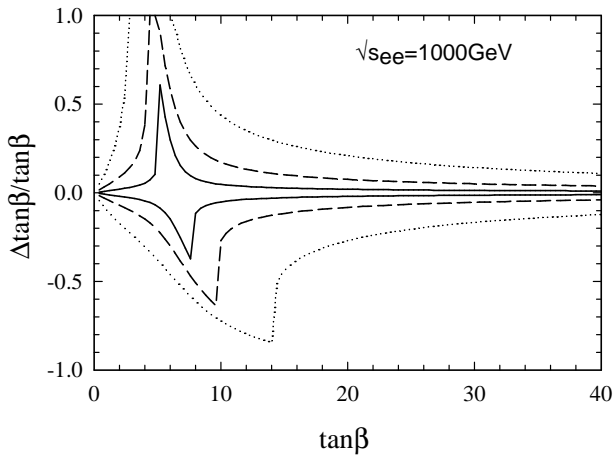,width=3.2in,clip=}}
\end{center}
\caption{$\Delta\tan\beta/\tan\beta$ {\it vs.} $\tan\beta$ for
the $\gamma\gamma$ (solid lines), $e\gamma$ (dashed lines)
and $e^+e^-$ (dotted lines) collider modes for $\sqrt{s}_{ee}=1$~TeV and 
$M_H = 200$~GeV.
}
\label{Fig5}
\end{figure}

As the charged Higgs boson has not yet been observed, its mass is not 
determined.  We expect that the charged Higgs boson will have been 
discovered and its mass determined by the time this analysis is performed by 
experimentalists, but for now we allow $M_H$ to vary.  In Fig.~6, we present 
results on the variation of $\Delta\tan\beta/\tan\beta$ with $M_H$ for 
$\sqrt{s}_{ee}=500$~GeV and $1$~TeV linear colliders operating in 
$\gamma \gamma$ mode.  The cross section decreases as we near the kinematic 
limit ($(M_H + m_t) \sim 0.8 \times \sqrt{s}_{ee}$) as does the sensitivity to the 
value of $\tan\beta$.  

\begin{figure}
\begin{center}
\centerline{\epsfig{file=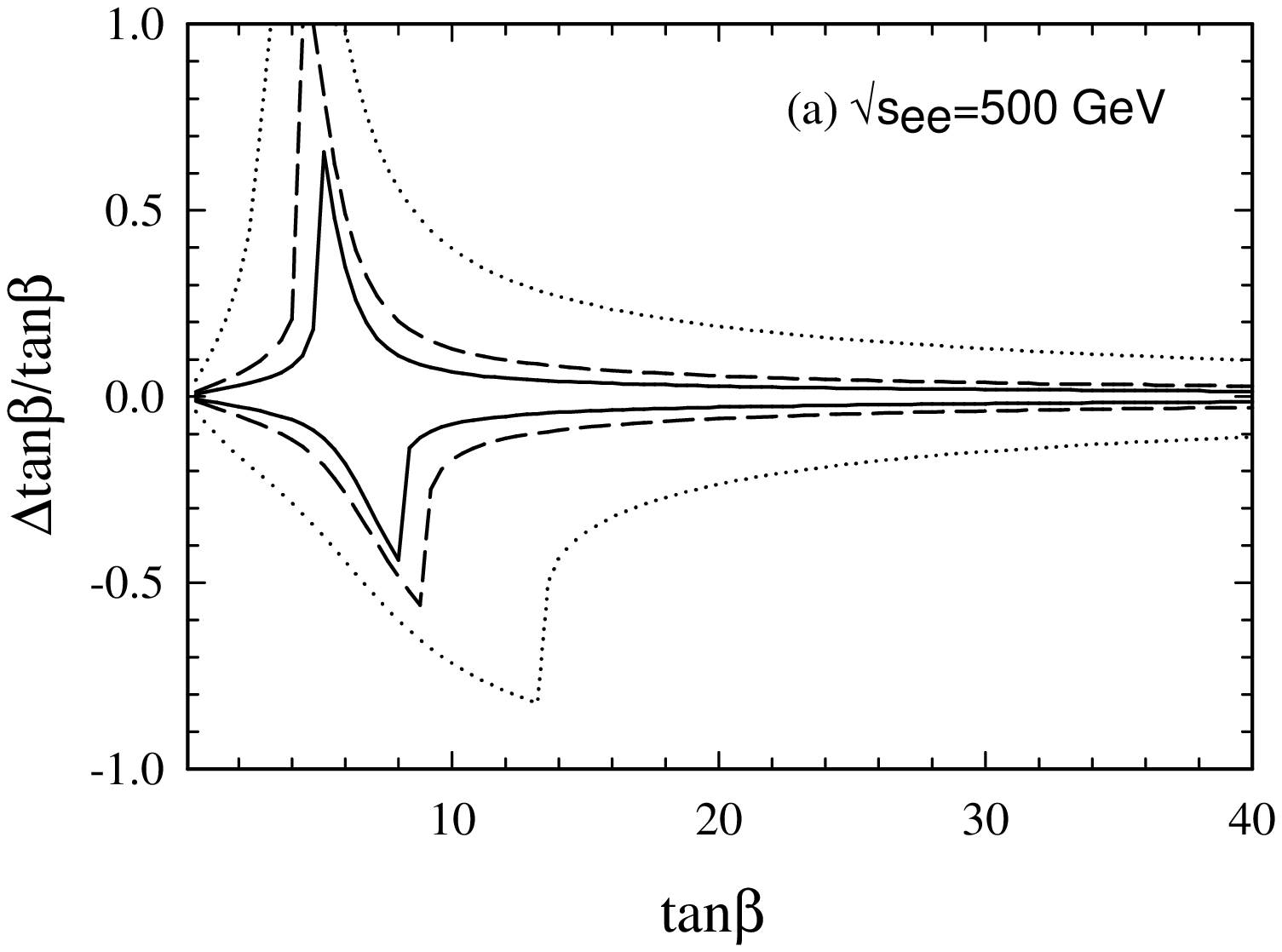,width=3.2in,clip=}}
\end{center}
\begin{center}
\centerline{\epsfig{file=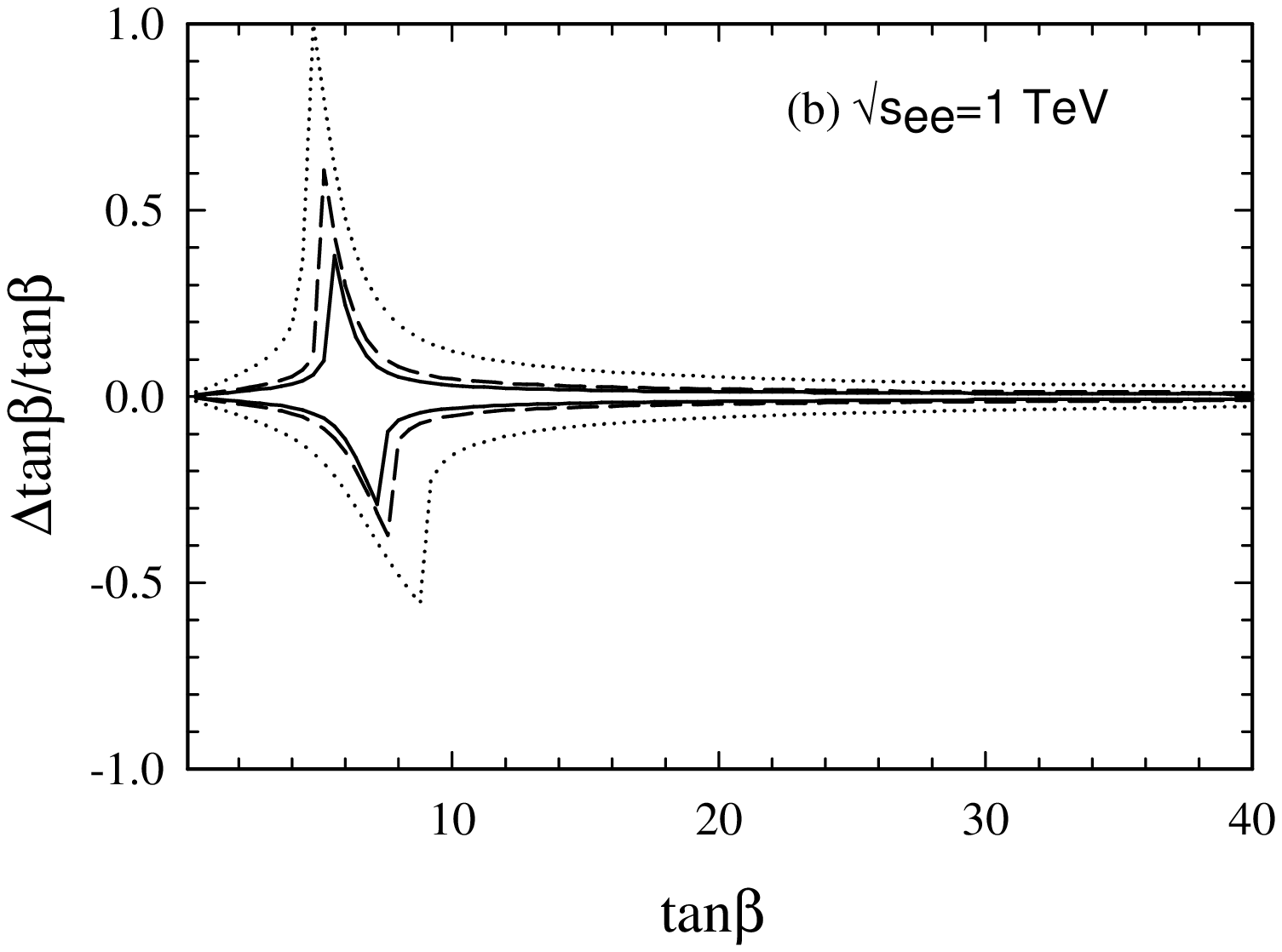,width=3.2in,clip=}}
\end{center}
\caption{Sensitivity to $\tan\beta$ for 
(a) a $\sqrt{s}_{ee}=500$~GeV $\gamma\gamma$ collider with
$M_H=100$~GeV (solid line),
150~GeV (dashed line) and 200~GeV (dotted line). 
(b) a $\sqrt{s}_{ee}=1$~TeV $\gamma\gamma$ collider with
$M_H=100$~GeV (solid line),
200~GeV (dashed line) and 400~GeV (dotted line). 
}
\label{Fig6}
\end{figure}

The central issue in making a precision measurement of $\tan\beta$ is 
the reconstruction efficiency of the signal.  Given that this depends 
on the details of the detector and can only be accurately estimated 
by performing a detailed detector Monte Carlo simulation in Fig 7
we show a 
series of results for reconstruction efficiencies of 100\%, 75\%, 
50\%, 25\%, 10\%, and 5\% for $\sqrt{s}_{ee}=500$~GeV (Fig 7a), 
800~GeV (Fig 7b), 1000~GeV (Fig 7c) and 1500~GeV (Fig 7d).  In this 
way, when the details of the detector are better known along with a 
good estimate of the reconstruction efficiency, one can use these 
figures to estimate the expected measurement error of $\tan\beta$ and 
compare the estimate from associated $t-H^{\pm}$ production to other 
processes considered in the literature.

\begin{figure*}
\begin{center}
\begin{minipage}[t]{7.0cm}
\centerline{\epsfig{file=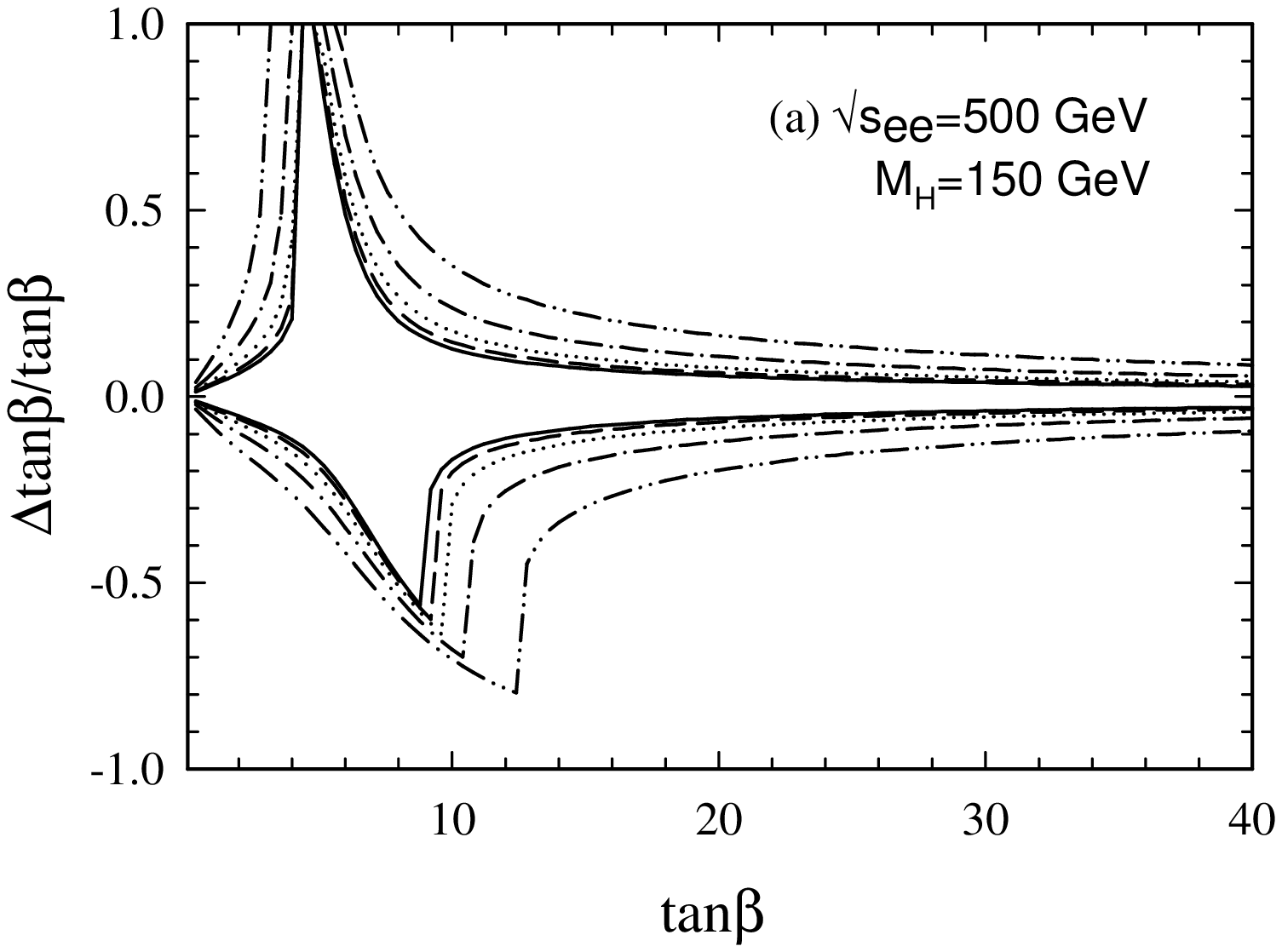,width=6.5cm,clip=}}
\end{minipage} \
\begin{minipage}[t]{7.0cm}
\centerline{\epsfig{file=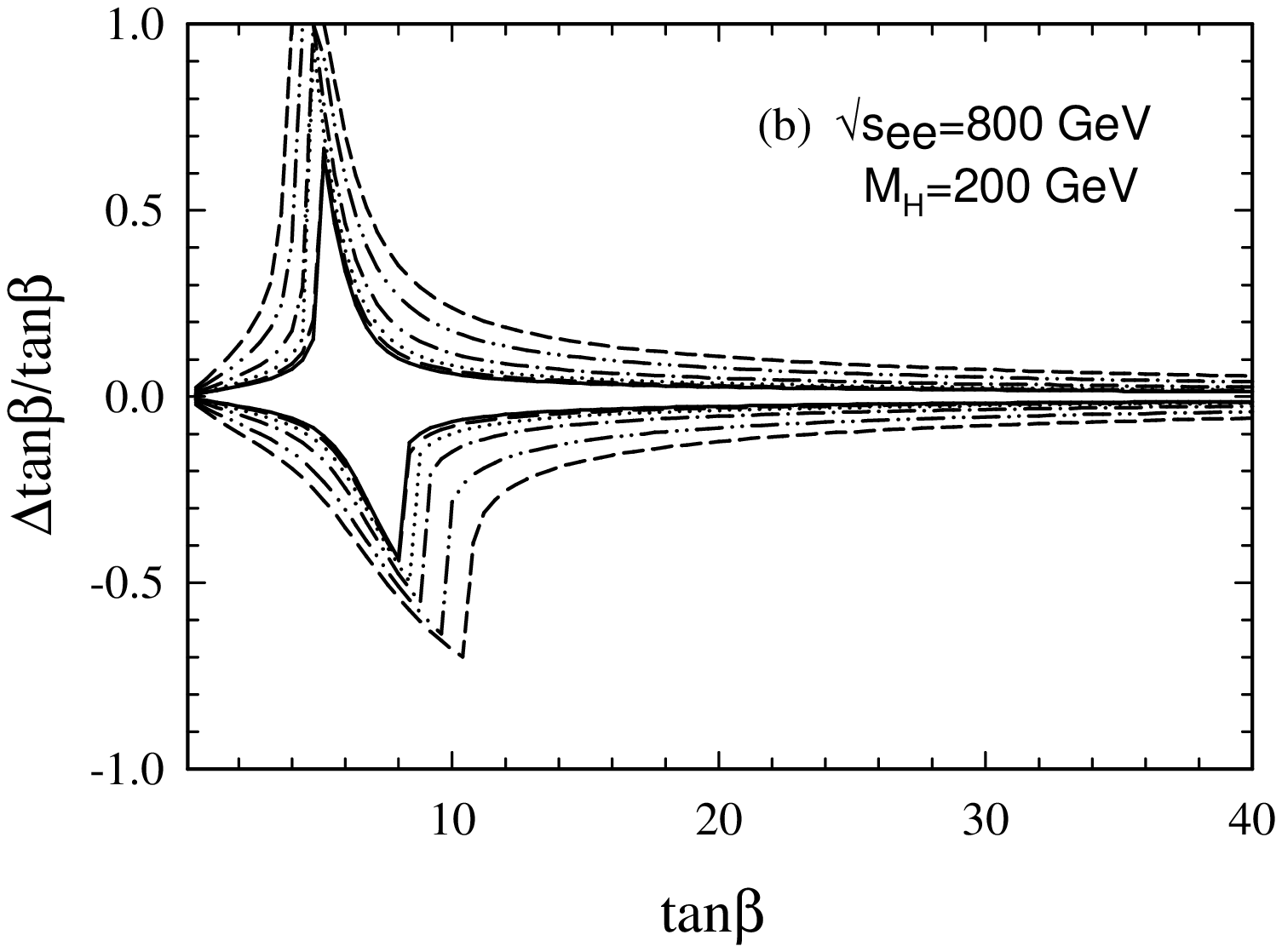,width=6.5cm,clip=} }
\end{minipage} \
\vskip 0.2cm
\begin{minipage}[t]{7.0cm}
\centerline{\epsfig{file=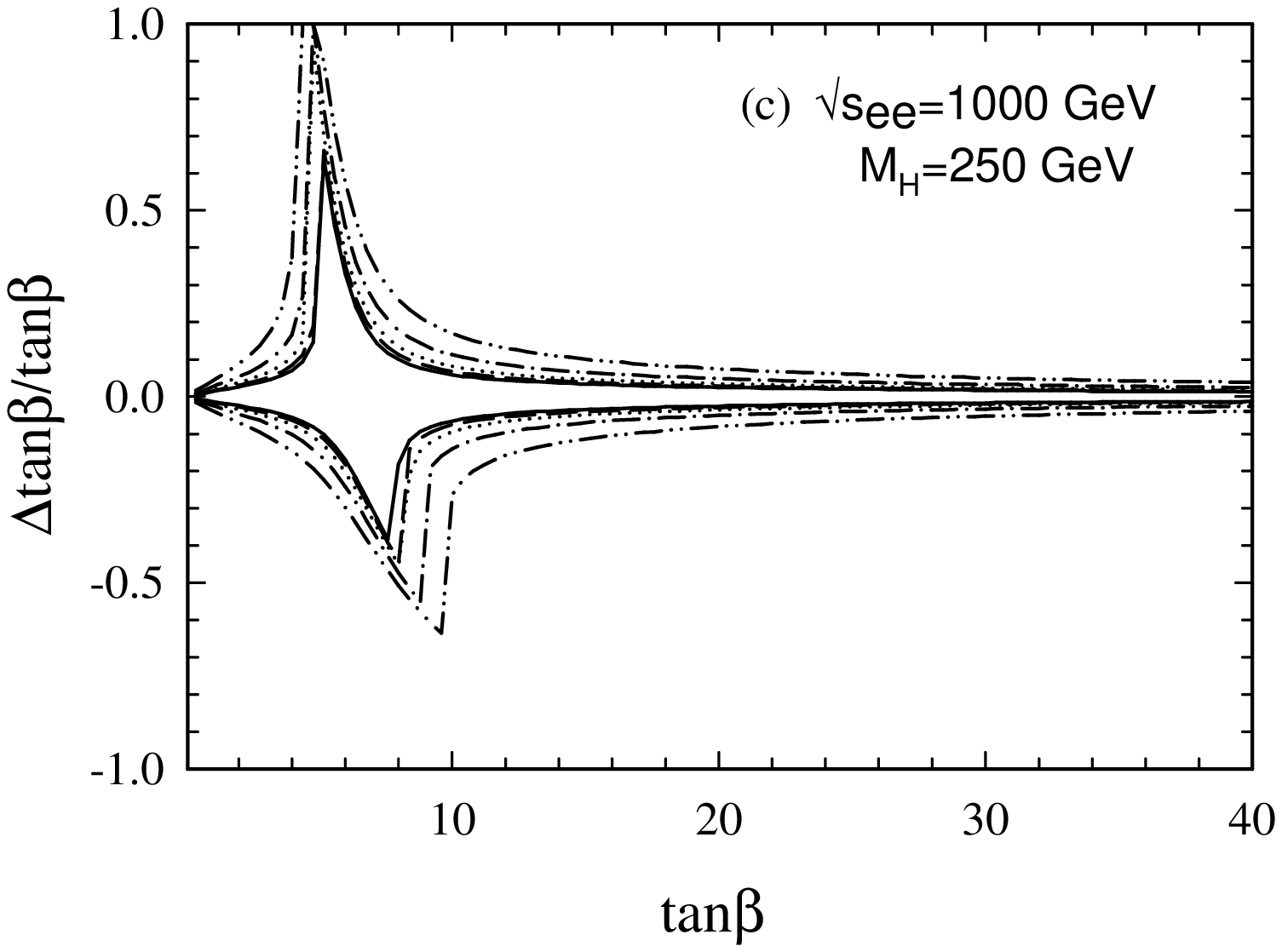,width=6.5cm,clip=}}
\end{minipage} \
\begin{minipage}[t]{7.0cm}
\centerline{\epsfig{file=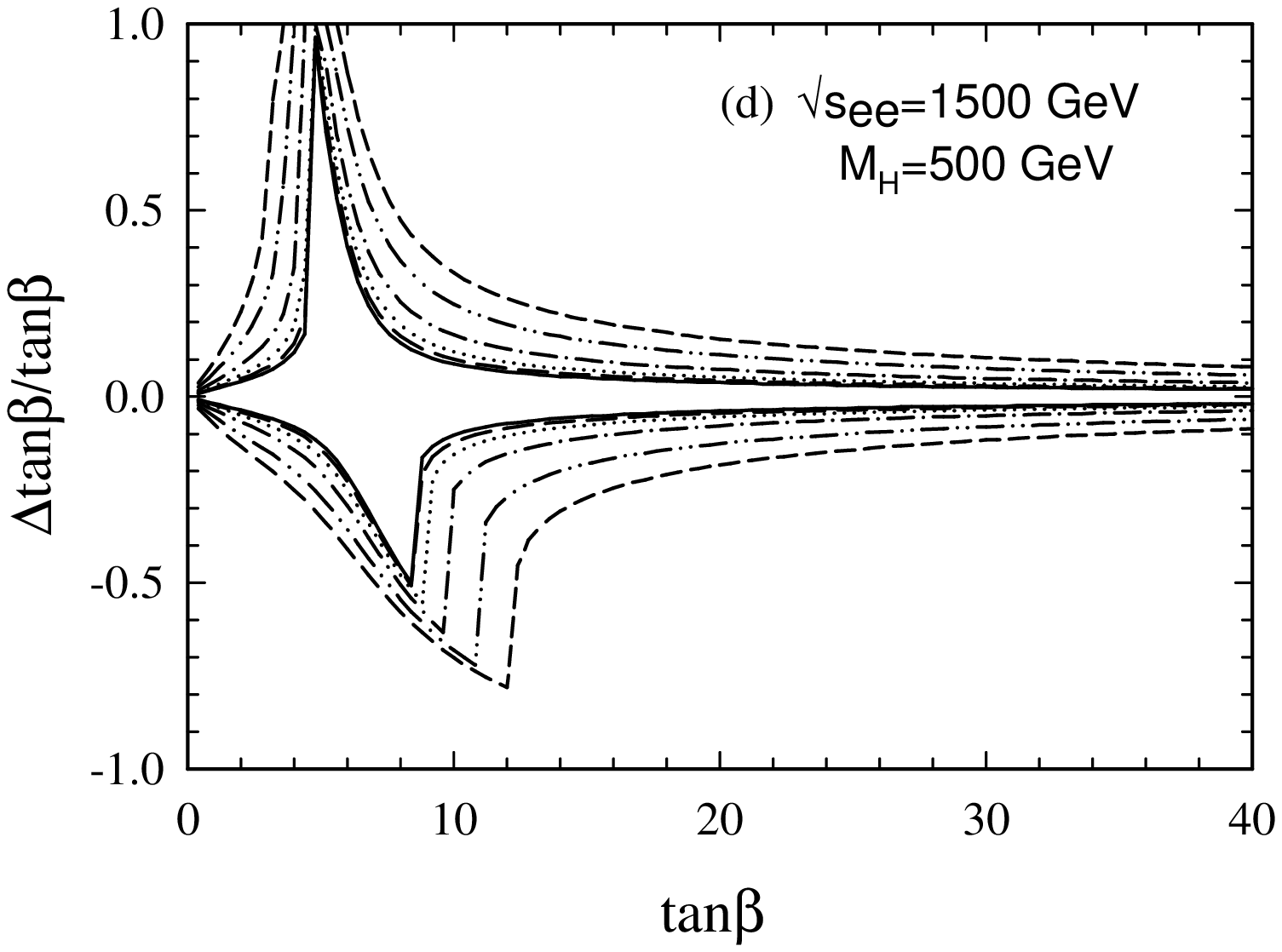,width=6.5cm,clip=} }
\end{minipage} \
\vspace{20pt}
\end{center}
\caption{Sensitivity to $\tan\beta$ assuming different reconstruction 
efficiencies for the $t-H^\pm$ final state.  
$\epsilon=100\%$ (solid line), 75\% (dashed line), 
50\% (dotted line), 25\% (dot-dashed line), 10\% (dot-dot-dashed line),
5\% (short-dashed line).  
Fig. (a) is for $\sqrt{s}_{ee}=500$~GeV, $M_H=150$~GeV;
(b) is for $\sqrt{s}_{ee}=800$~GeV, $M_H=200$~GeV;
(c) is for $\sqrt{s}_{ee}=1000$~GeV, $M_H=250$~GeV;
and (d) is for $\sqrt{s}_{ee}=1500$~GeV, $M_H=500$~GeV.
}
\label{Fig7}
\end{figure*}

\section{Conclusions}

In this paper we studied the potential for measuring the parameter $\tan\beta$ 
arising in type II Higgs doublet models such as MSSM using associated 
$t$-quark charged Higgs boson production in $\gamma\gamma$ collision; 
$\sigma(\gamma\gamma \to t H^\pm +X)$.  We find that sensitivity to the value 
of $\tan\beta$ can be as small as several percent or as large ${\cal O}
(100\%)$, 
depending on the value of $\tan\beta$.  For this process, the region of 
$\tan\beta \simeq 7$ is particularly insensitive to the value of $\tan\beta$.  
Overall, this process is competitive with those considered by Feng and 
Moroi~\cite{feng97}, Barger~{\it et al.,}~\cite{barger01} and 
Gunion~{\it et al}~\cite{gunion02}.  Feng and Moroi consider the production of 
two Higgs bosons, $HA$ and $H^+H^-$, as well as associated $H^- t \bar{b}$ and 
find good sensitivity at low $\tan\beta$.  Barger {\it et al} consider 
associated production of heavy neutral Higgs and heavy quark pairs, 
$Hb\bar{b}$, $Ht\bar{t}$, $Ab\bar{b}$ and $At\bar{t}$, and also find good 
sensitivity at low $\tan\beta$.  Gunion {\it et al} also consider 
Higgs boson associated production with heavy quark pairs, as well as the 
production of two Higgs bosons, $HA$ and $H^+H^-$, with four heavy quarks in 
the final state.  Their combined analysis indicates that 
$\Delta\tan\beta/\tan\beta \sim$~a few to ten percent is possible for 
$\tan\beta>30$ and $\tan\beta<10$, but there remains a potential hole for 
intermediate values of $\tan\beta$: $10<\tan\beta<30$.  The analysis 
considered here provides sensitivity to very low $\tan\beta \leqnew$~3 and for 
intermediate to large values of $\tan\beta \geqnew$~10.  Thus, it should be 
considered an additional tool in disentangling the Higgs sector of the 
electroweak theory and complements other processes previously considered.

\acknowledgments

SG thanks Chris Hearty and Peter Zerwas for useful discussions.  MAD thanks 
Zack Sullivan for useful discussions. 
This research was supported in part by the Natural Sciences and Engineering 
Research Council of Canada.  The work of M.A.D.\ was supported, in part, by 
the Commonwealth College of The Pennsylvania State University under a Research 
Development Grant (RDG).

\end{document}